# Sifaka: Text Mining Above a Search API

Cameron VandenBerg
Language Technologies Institute
Carnegie Mellon University
Pittsburgh, PA 15213, USA
cmw2@cs.cmu.edu

Jamie Callan
Language Technologies Institute
Carnegie Mellon University
Pittsburgh, PA 15213, USA
callan@cs.cmu.edu

## ABSTRACT

Text mining and analytics software has become popular, but little attention has been paid to the software architectures of such systems. Often they are built 'from scratch' using special-purpose software and data structures, which increases their cost and complexity.

This demo paper describes `Sifaka`, a new open-source text mining application constructed above a standard search engine index using existing application programmer interface (API) calls. Indexing integrates popular annotation software libraries to augment the full-text index with noun phrase and named-entities; n-grams are also provided. `Sifaka` enables a person to quickly explore and analyze large text collections using search, frequency analysis, and co-occurrence analysis; and import existing document labels or interactively construct document sets that are positive or negative examples of new concepts, perform feature selection, and export feature vectors compatible with popular machine learning software. `Sifaka` demonstrates that search engines are good platforms for text mining applications while also making common IR text mining capabilities accessible to researchers in disciplines where programming skills are less common.

## 1. INTRODUCTION

Text mining is important to industry, governments, researchers, and educators, however there is little good open-source text mining software. This statement may sound surprising because there are widely-available machine learning toolkits (e.g., `Weka [1]`, `Mallet [2]`, `SVMLight [3]`, `scikit-learn [4]`) and text analysis components (e.g., from the Stanford NLP group [5, 6] and LingPipe [7]). However, running hundreds-of-thousands or millions of documents through a named-entity annotator or part-of-speech tagger; stemming tokens; weighting and selecting features; and producing feature vectors for a machine learning toolkit requires programming skills and some expertise. Ph.D.-level researchers consider these tasks routine, but they are an obstacle to people with fewer information retrieval and computer science skills.

Usually text mining software is viewed as distinct from search software, however we argue that search engines are a natural foundation for text mining systems. Full-text search engines are efficient and scalable language databases that ingest material in multiple formats, support text attributes and annotations, and provide powerful query languages for defining new concepts and locating desired information. Typically their application programming interfaces (APIs) support high-level queries as well as access to low-level statistics, information about what occurs where, and parsed representations of each document. Building text mining software above a good search engine allows a software developer to inherit those capabilities and focus effort on capabilities unique to the text mining application.

This demo paper presents `Sifaka`, a new open-source text mining tool that tests this perspective. `Sifaka` was developed as a proof-of-concept, to explore the limits of what can be accomplished within a standard search engine API, and to provide several common text mining capabilities to text analysts, researchers, and students in related fields (e.g., information science, public policy, and social science). `Sifaka` was built above the `Lucene [8]` search engine's application programming interface and index; however the functionality and data structures that it uses are commonly available in today's search engines.

The next section explores the requirements that a text mining system places on its language database. Section 3 describes `Sifaka`'s architecture, how it is supported by Lucene, and how it could be supported by another search engine such as `Galago [9]`. Section 4 provides a set of case studies that illustrate `Sifaka`'s text mining capabilities. Finally, Section 5 concludes.

## 2. REQUIREMENTS

*Text mining* is a general, loosely-defined phrase that covers many forms of text analysis; no system could cover them all. However, some general components arise frequently:

- Lexical or linguistic analysis to derive features;
- Selection of documents that satisfy some criterion or pattern;
- Frequency and co-occurrence analysis of concepts; and
- Use of text categorization.

The next section discusses how `Sifaka` uses `Lucene` to provide these capabilities in a convenient software application.

## 3. SOFTWARE ARCHITECTURE

`Sifaka` consists of two software applications: An application that constructs a search engine index, and an application that supports a variety of search and text mining capabilities. Both applications are implemented using `Lucene` search engine libraries.

### 3.1 Indexing

The indexing application supports several document formats, for example, .txt, HTML, some TREC, and Twitter JSON formats. Some formats support embedded metadata within documents, for example category labels. Parsers transform documents in specific formats into generic structured document objects that contain metadata and fields, as is typical in `Lucene`.

The document object is enhanced by optionally passing each text field through an extensible sequence of text analysis annotators that can create new document fields. A text analysis annotator consists of an analysis component and a lightweight wrapper that converts the analysis output into a sequence of index terms. For example, a named entity annotator returns an annotated sequence (e.g., "… Angela/PERSON Merkel/PERSON …") that its wrapper converts to a sequence of index terms (e.g., "… angela_merkel …"). Words that are discarded by a wrapper are replaced by a stopword. For example, "The president is Donald Trump" is transformed to "a a a donald_trump a." The stopwords cause Lucene to increment its location counter during indexing, thus preserving distance relationships and maintaining approximate alignment between terms produced by text analysis plug-ins and terms in the original text. Text analysis annotators store their results in one or more new document fields. For example, the named entity annotator stores results in PERSON, LOCATION, and ORGANIZATION fields.

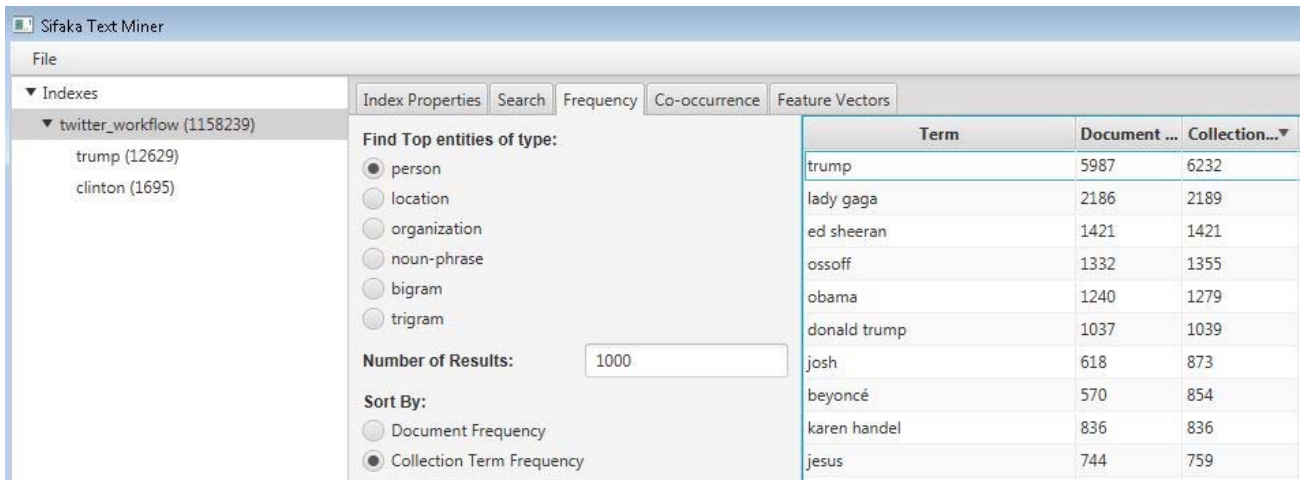

**Figure 1:** Frequency analysis of the Twitter 1% Spritzer feed on June 21, 2017

Three text analysis annotators are provided by default. One annotator adds bigrams and trigrams (e.g., 'fat_cat', 'third_party_candidate') to the document representation. Two annotators add named entities and noun phrases identified by the popular Stanford named entity [6] and part of speech [5] taggers. New annotators can be added easily.

Lucene's standard indexing classes use the structured document object to create several types of index data structures, for example inverted index, forward index[1], and string lookup data structures.

### 3.2 Text Mining

`Sifaka` supports a variety of search and text mining capabilities. This section provides a high-level overview of those capabilities, with a focus on how they are supported by a standard search engine index and API. From an architectural perspective, `Sifaka` can be thought of as having four main components that can be combined in different ways to support interactive text mining.

**Search:** `Sifaka` uses `Lucene`'s search capabilities, including its structured query language, to provide interactive search. Search results can be saved as named *saved sets*. Figure 1 shows the work of a person that created two saved sets. A saved set named 'trump' contains 12,629 documents; another named 'clinton' contains 1,695 documents. Saved sets are stored in memory and can be combined to form new saved sets. Any search or text mining action that can be performed on a `Lucene` index can also be performed on a saved set.

**Frequency analysis** is supported for all field types, for example, title terms, body terms, noun phrases, and different types of named entities. The result is a list of field-specific index terms sorted by document frequency or collection term frequency. `Sifaka` uses `Lucene`'s forward index to build a list of entities and their frequencies.

**Co-occurrence analysis** is supported for all field types using two common co-occurrence metrics (pointwise mutual information and phi-square [10]). Most co-occurrence metrics require filling a contingency table for each pair. Queries along with the forward index support this operation. For example, to find all *companies* that co-occur with Donald Trump, a query retrieves a list of documents containing Donald Trump (*count* ($x$)). The forward index entries of the returned documents (or optionally just the top n) are examined to find and generate counts for co-occurring entities (*count* ($x$ AND $y$)) and *count* ($x$ AND $\overline{y}$)). Entities that co-occur less often than a user-defined threshold may be eliminated. Statistics for each co-occurring entity are fetched from the term dictionary to fill the contingency table (*count* ($\overline{x}$ AND $y$) and *count* ($\overline{x}$ and $\overline{y}$)) and calculate the co-occurrence metric. Finally, the list is sorted into descending order and displayed.

**Feature Vectors** may be created for a set of documents defined by a metadata value (e.g., a category label), a search, or a saved set. These documents are treated as positive examples of a concept. Documents not in the saved set – either all of them or a random sample – are treated as negative examples. Features may be drawn from any document field, for example, title unigram, body bigram, and body person. Cohen's kappa indicates the strength of association between each feature and the set of documents. Kappa values are calculated for each feature and category using a process and data structures similar to calculating co-occurrence values. Features and their kappa values for each category are returned in descending order. Feature selection may be performed based on the number of features and/or a minimum kappa threshold. Feature weights may be selected to be binary, term frequency (tf), or term frequency with inverse term frequency (tf.idf) weights. These choices are used to prune copies of the forward index entries of each selected document and export them as feature vectors in

---

[1] A *forward index* provides rapid access to a list of the terms that occur in a document. It may also store the location of each term.

| Term | Term Freq | PMI ▼ | Phi-Square |
|---|---|---|---|
| win congrats karen handel | 7 | 7.20897 | 0.00816 |
| cnn cuts | 8 | 7.20897 | 0.00933 |
| congrats karen handel | 30 | 7.20897 | 0.03498 |
| #trump endorsement | 7 | 7.20897 | 0.00816 |
| karen handel | 647 | 7.20897 | 0.75482 |
| woman rep ga | 102 | 7.20897 | 0.11894 |
| karen handel victory party | 10 | 7.20897 | 0.01166 |
| karen handel thanks president tru... | 5 | 7.20897 | 0.00583 |
| republican karen handel | 21 | 7.20897 | 0.02449 |
| karen handel thanks potus | 26 | 7.20897 | 0.03032 |
| glass ceiling stories | 103 | 7.20897 | 0.12011 |
| trump triumphs | 5 | 7.20897 | 0.00583 |

**Figure 2:** Noun phrases that have the highest co-occurrence with the person *Karen Handel* in the Twitter 1% Spritzer feed on June 21. 2017.

Weka's ARFF format [1]. The options available to creating feature vectors are displayed in Figure 4.

### 3.3 Other Search Engines

Little of `Sifaka` is specific to `Lucene`. Most recent search engines provide functionality and APIs similar to the search engine functionality and APIs that `Sifaka` requires. `Sifaka` is written in Java, thus it would be easiest to port it to other search engines written in Java. We looked carefully at what would be required to port `Sifaka` from `Lucene` to `Galago` [9], an open-source search engine used by the research community. `Galago` provides a lexical analysis pipeline that would accommodate Sifaka's text analysis plug-ins; comparable search capabilities; and access to term dictionary and inverted list data structures. Older versions of `Galago` do not provide a forward index, but that capability was added recently in version 3.12.

## 4. TEXT MINING WITH SIFAKA

`Sifaka`'s capabilities are illustrated below by four examples.

### 4.1 Frequency Analysis: A Day of Twitter

Analysis of a new and unknown corpus or information stream often begins with frequency analysis to discover the concepts that are the greatest focus of discussion.

On any day, Twitter's 1% *Spritzer* microblog feed contains millions of tweets. A person can use `Sifaka` to quickly determine the most actively discussed concepts of a particular type on a particular day. For example, on June 21st, 2017, the person name *Beyoncé* occurs frequently (Figure 1). Although Beyoncé is a well-known celebrity, most days she is not one of the 20 most discussed entities. Clicking on her name switches to the search tab. A quick search on her name retrieves tweets that discuss the pictures with her new twins that had just been published, which explains the frequency on that day.

### 4.2 Co-Occurrence Analysis: Digging Deeper

Frequency analysis may reveal unfamiliar names, for example, the name *Karen Handel* in Figure 1. A search on the name and examination of a few documents may not provide sufficient context to understand who the person is or why they are discussed so frequently. Co-occurrence analysis may provide more information. For example, a few of the noun-phrases that have the strongest pointwise mutual information (PMI) with *Karen Handel* are shown in Figure 2. It includes phrases such as "win congrats Karen Handel", "#trump endorsement", "woman rep ga", and "republican Karen Handel", which suggests that she is a politician in the Republican party from the state of Georgia (GA) in the United States. In fact, she was a republican candidate who won the Georgia state special election the day before.

| Term | Term Freq | PMI | Phi-Square ▼ |
|---|---|---|---|
| john podesta | 53 | 7.9391 | 0.1283 |
| barack obama | 53 | 7.18343 | 0.06022 |
| tom cotton | 14 | 7.4028 | 0.01981 |
| jake tapper | 7 | 7.55695 | 0.01156 |
| julius caesar | 7 | 6.45833 | 0.00385 |
| kamala harris | 6 | 6.15003 | 0.00242 |
| nancy pelosi | 11 | 5.11394 | 0.00156 |
| mueller | 8 | 4.38849 | 5.4E-4 |
| karen handel | 9 | 3.56454 | 2.6E-4 |
| trump | 12 | 1.8835 | 5.0E-5 |

**Figure 3:** People that have the highest co-occurrence with the person *Hillary Clinton* in the Twitter 1% Spritzer feed on June 21, 2017.

Frequency analysis may also reveal terms that are familiar and *often* frequent, for example the phrase "Hillary Clinton". Co-occurrence analysis provides greater context about the discussion of these entities, too. A person co-occurrence analysis using the phi-square co-occurrence metric for June 21st, 2017 shows that John Podesta, Barack Obama, and Tom Cotton co-occur highly with Hillary Clinton on that day (Figure 3). Searching for John Podesta reveals that there were tweets about John Podesta accepting money from Russia while advising Hillary Clinton and Barack Obama. An analysis of co-occurring organizations on the same day returns the State Department as the highest co-occurring organization. Searching for the State Department reveals that there was discussion about the State Department revoking Hillary Clinton's security clearance. These analyses show the issues surrounding Hillary Clinton that day.

### 4.3 Training Classifiers: Existing Categories

It is common to use a labeled corpus to train a text classifier that can be used on a similar corpus. For example, one might use a set of Reuters news documents that are labeled with topic, industry, and region categories to train classifiers that can recognize news documents about topics such as the financial industry or regions such as Europe. Good machine learning toolkits such as `Weka` [1] are freely available. However, transforming a text corpus into feature vectors can be an obstacle for people with few or no programming skills. `Sifaka` supports this activity.

To create feature vectors, a user can go to the feature vector tab and select document labels such as earn, acq (acquisitions), money-fx, and interest as shown in Figure 4. `Sifaka` calculates the kappa values for the features in each set and for the documents that do not fall into any of those sets if desired. Once the kappa values are calculated, the user can choose how many features and/or the kappa cutoff for each category and create a feature vector file in ARFF format, which can be used with WEKA.

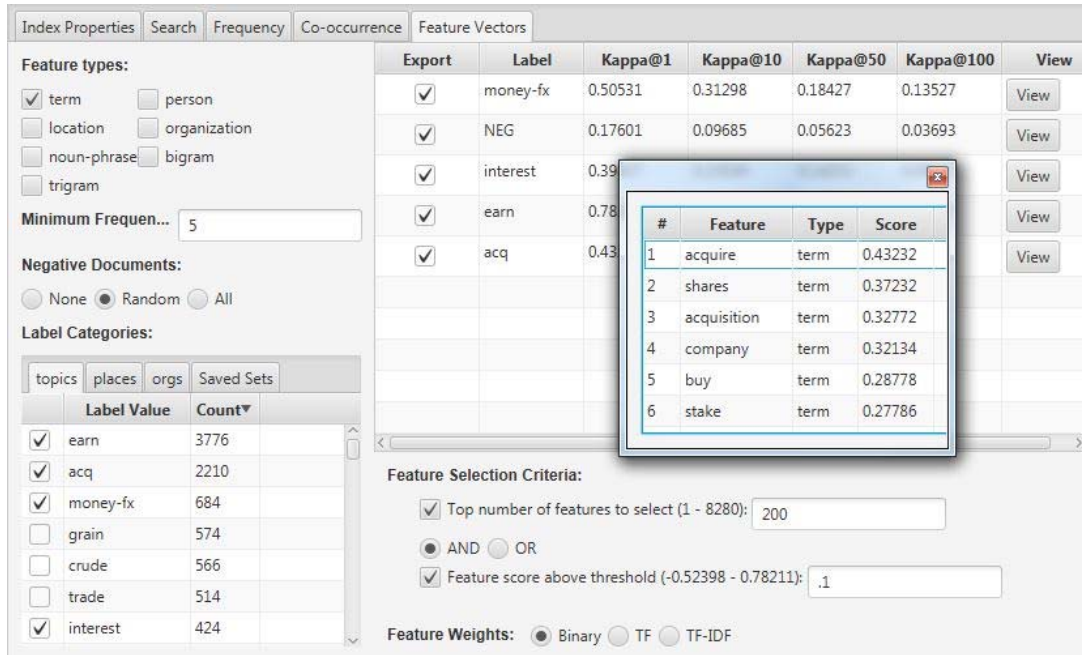

**Figure 4:** Feature selection on the Reuters-21578 dataset.

### 4.4 Training Classifiers: New Categories

A researcher studying the U.S. 2016 presidential election might want to find news documents about that topic. However, there may not be labeled documents for training a classifier about this topic. This problem is easily solved using Sifaka's search capabilities.

Words, phrases, and named-entity references can be used to create a query that finds documents about a presidential candidate such as Hillary Clinton. The top-ranked results are used to define a new saved set (e.g., 'clinton' in Figure 2). This process is repeated for other presidential candidates, for example, Donald Trump, Ted Cruz, and Bernie Sanders. Finally, the saved sets are combined to form a set of positive examples. Negative examples are defined using a similar process, or they can be selected randomly from the corpus.

Once positive and negative examples are defined, feature selection and export of feature vectors is done as defined above (Section 4.3).

### 5. CONCUSION

This paper and its accompanying demo present Sifaka, a new open-source text mining application built above the Lucene search engine. Sifaka demonstrates that useful text mining software can be developed using standard search engine data structures and APIs, while also providing several common text mining capabilities to text analysts, researchers, and students in fields that rely on text analysis.

Sifaka's initial functionality demonstrates a range of activities that can be accomplished using common text analysis, structured documents, search, search engine data structures, and feature vectors. Other common text mining functionality fits within the framework defined by Sifaka and common search engine APIs. For example, clustering for arbitrary sets of documents can be done using feature vectors generated from forward index entries, as is done for text classification.

We hope that Sifaka will encourage greater use and study of search engine indexes and APIs as *language databases* capable of supporting diverse text analysis applications.

### 6. ACKNOWLEDGEMENTS

This research was supported by National Science Foundation grant CNS-1405045. Any opinions, findings, and conclusions in this paper are the authors' and do not necessarily reflect those of the sponsor.